\documentclass[vecphys]{svmult}
\usepackage{graphicx} 
\usepackage{txfonts} 
\usepackage{natbib} 
\usepackage[figuresright]{rotating} 
\usepackage{subfigure} 
\usepackage{lscape} 
\usepackage{psfig} 
\usepackage{color} 

\usepackage{makeidx}         
\usepackage{graphicx}        
\usepackage{multicol}        
\usepackage[bottom]{footmisc}

\makeindex             

\begin{document}

\title*{HII Galaxies with 3D Spectroscopy}
\authorrunning{Kehrig et al.}
\author{C. Kehrig\inst{1,2},
J.M. V\'{\i}lchez\inst{1}, S.F. Sanchez\inst{3}, E. Telles\inst{2}, D. Martin-Gord\'on\inst{1},
\and L. L\'opez-Mart\'{\i}n\inst{4}}
\institute{Instituto de Astrof\'{\i}sica de Andaluc\'{\i}a (CSIC), Camino Bajo de Huetor, Apartado 3004, 18080 Granada, Spain 
\texttt{kehrig@iaa.es,jvm@iaa.es,dmg@iaa.es}
\and Observat\'{o}rio Nacional, Rua Jos\'{e} Cristino, 77, E-20.921-400, Rio de Janeiro, Brazil 
\texttt{kehrig@on.br,etelles@on.br}
\and Calar Alto Observatory (CAHA), Almer\'{\i}a E-040004, Spain 
\texttt{sanchez@caha.es}
\and Instituto de Astrof\'{\i}sica de Canarias, c/ V\'{\i}a Láctea s/n, 38200, La Laguna, Tenerife, Spain \texttt{luislm@ll.iac.es}}
%
%
\maketitle
\index{C. Kehrig}

\begin{abstract}

We present preliminary results from our ongoing work on integral field
spectroscopy (IFS) of the three prototypical HII galaxies IIZw70,
IIZw71 and IZw18 \cite{P02}. The observations were taken with the
instruments INTEGRAL/WYFFOS at the WHT (ORM, La Palma) and PMAS (Calar
Alto, Almeria), covering a spectral range from 3600 to 6800 \AA~thus
including from [OII]$\lambda$3727 $\AA$ up to the
[SII]$\lambda$$\lambda$6717,31 $\AA$ lines.  Two-dimensional
spectroscopy allows us to collect simultaneously the spectra of many
different regions of an extended object, combining photometry and
spectroscopy in the same data set. The great advantage of using IFS
for the investigation of galaxies is that it allows us to obtain data
on the galaxy positions, velocity fields and star-formation properties
all in one data cube. Our main aim is to perform a bidimensional study
of the ionization structure, physical-chemical conditions and the
velocity field of the ionized gas in these galaxies \cite{V98}. Maps
of the relevant emission lines and a preliminary analysis of the
ionization structure are presented for IIZw70.

\end{abstract}

\section{Introduction}
\label{sec:1}

HII galaxies are low-metallicity objects
undergoing violent star formation \cite{SS}.Their
optical spectra show strong emission lines (recombination lines of
hydrogen and helium, as well as forbidden lines of elements like
oxygen, neon, nitrogen, sulfur, among others) that are very similar to
the spectra of extragalactic HII regions (e.g. \cite{T91}). Among
them we can find the least chemically-evolved galaxies in the local
Universe.Due to their proximity they offer ideal laboratories for
studying massive star formation with an accuracy and spatial
resolution that cannot be achieved in similar studies in the distant
Universe.

The Integral Field Unit (IFU) spectrographs allow simultaneous spectra
to be taken of extended objects.  From this, we can learn how the
interstellar medium (ISM) is structured, and derive physical
conditions and chemical properties corresponding to individual
star-forming knots and external ionized gas component (located at more
or less one kpc from the main body of the galaxy). The study of the
distribution of these properties is an important issue for our
understanding of the interplay between the massive stellar population
and the ISM, unveiling the mechanisms that govern star formation.
Studying the chemical abundance distribution of the ionized
gas can be used as a tool to explore the possible chemical evolution
paths.

Our sample is composed by three HII galaxies: IZw18, IIZw70 and
IIZw71.  IZw18, since pioneering work of \cite{SS}, has remained a
unique object for the study of the chemical evolution galaxies, given
its very low oxygen abundance, approximately 1/50 solar, and the
absence of a significant underlying old population. However, whether
this is a signature of youth remains a subject of debate
(e.g. \cite{P02}). The optical structure of this galaxy is very rich,
including the two (southeast and northwest) central bright knots
within the main body \cite{V98}. IIZw70 and IIZw71 are two prototypical
HII galaxies that are interacting. IIZw70 presents a conspicuous
central star-forming region, and extending to the north-west and
south-west there appear fan-like extensions, resembling jet structures
\cite{C01}. IIZw71 shows a chain of individual star-forming knots that
crosses the whole optical body of the galaxy and accounts for all the
H$\alpha$ emission.

\begin{figure}[!ht]
 \mbox{
  \centerline{
\hspace*{0.0cm}\subfigure{\label{rotcen-a}\includegraphics[width=3.3cm]{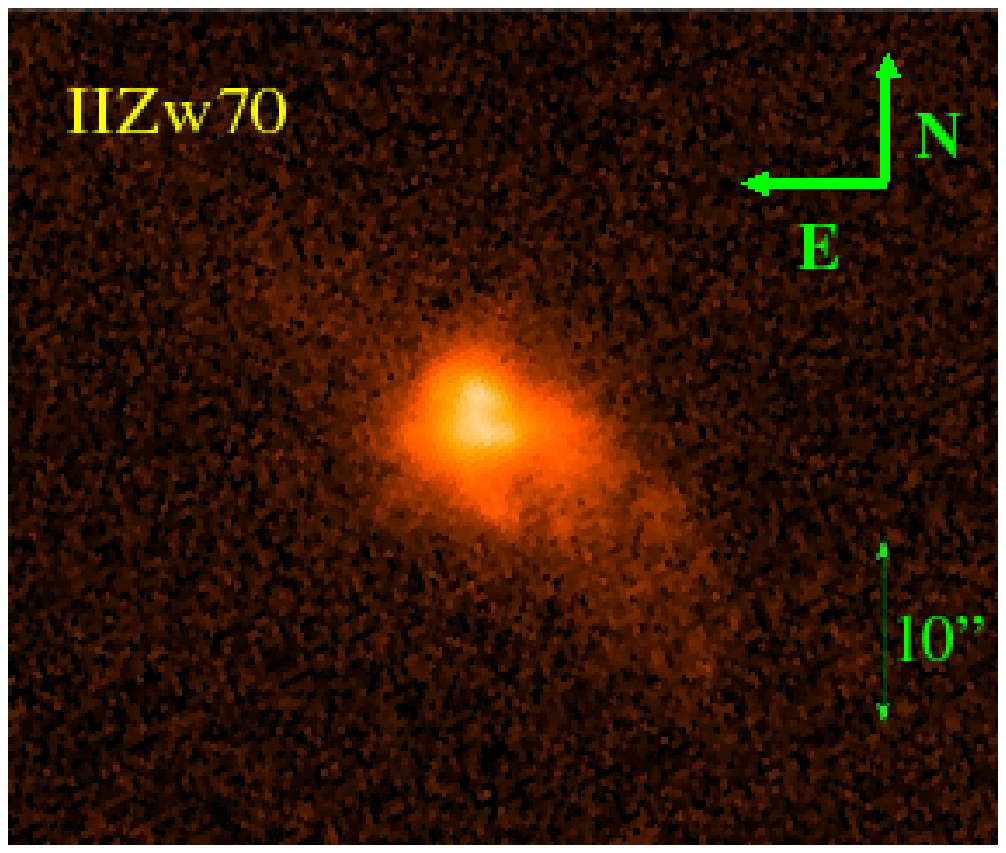}}
\hspace*{-0.04cm}\subfigure{\label{rotcen-a}\includegraphics[width=3.3cm]{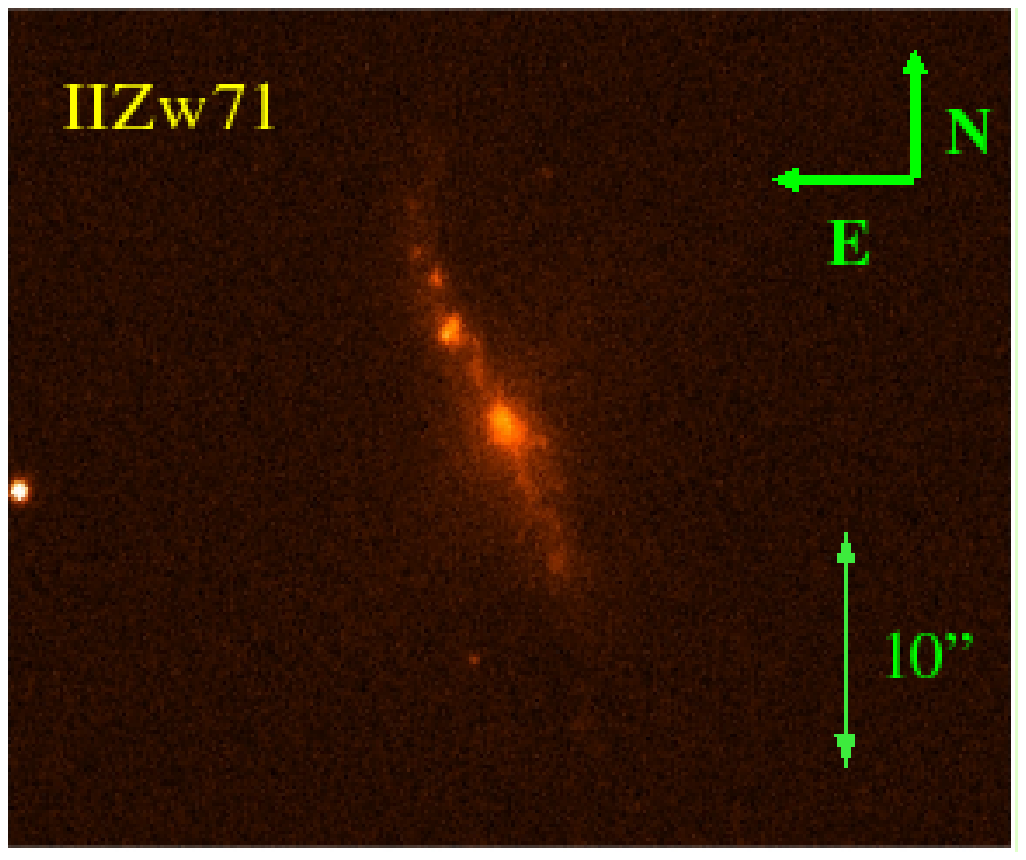}} 
\hspace*{-0.04cm}\subfigure{\label{rotcen-a}\includegraphics[width=3.3cm]{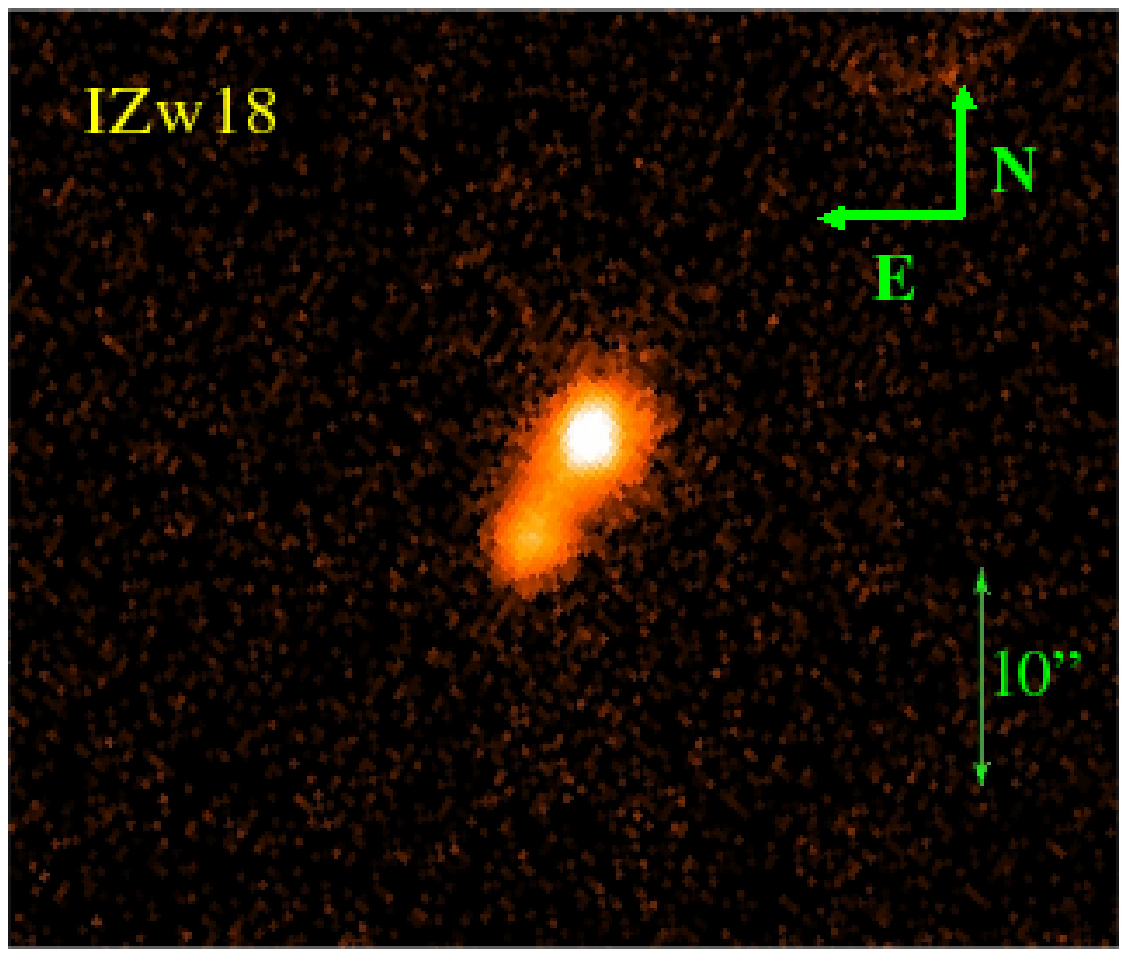}}
   }}\\
\caption
{Images of the HII galaxies IIZW70, IIZW71 and IZW18 in the U band
from Sloan Digital Sky Survey (The SDSS web site is
http://www.sdss.org). North is up and east is left.}

\label{sdss}
\end{figure}  

\section{Results}
\label{sec:2}

Two-dimensional spectra have been analyzed in order to derive the
properties of the different components of the galaxy IIZw70, that was
observed with PMAS at CAHA.

The PMAS spectrograph is equipped with 256 fibers coupled
to a $16\times16$ lens array. Each fiber has a spatial sampling of 0
$.\!\!^{\prime\prime}$5$~\times~$0 $.\!\!^{\prime\prime}$5 on the sky
resulting in a field of view of $8\hbox{$^{\prime\prime}$
}\times8\hbox{$^{\prime\prime}$ }$. Mosaic patterns were used to cover
the central regions and different regions of interest (e.g., tidal
tails). The data were reduced using the software R3D \cite{SA06}.
We fitted the line profiles to derive the integrated flux of each
emission line, using the software FIT3D \cite{S06}, already used in previous similar studies \cite{G05},\cite{S04}.

\begin{figure}[!ht]
 \mbox{
  \centerline{
\hspace*{0.0cm}\subfigure{\label{rotcen-a}\includegraphics[width=3.3cm]{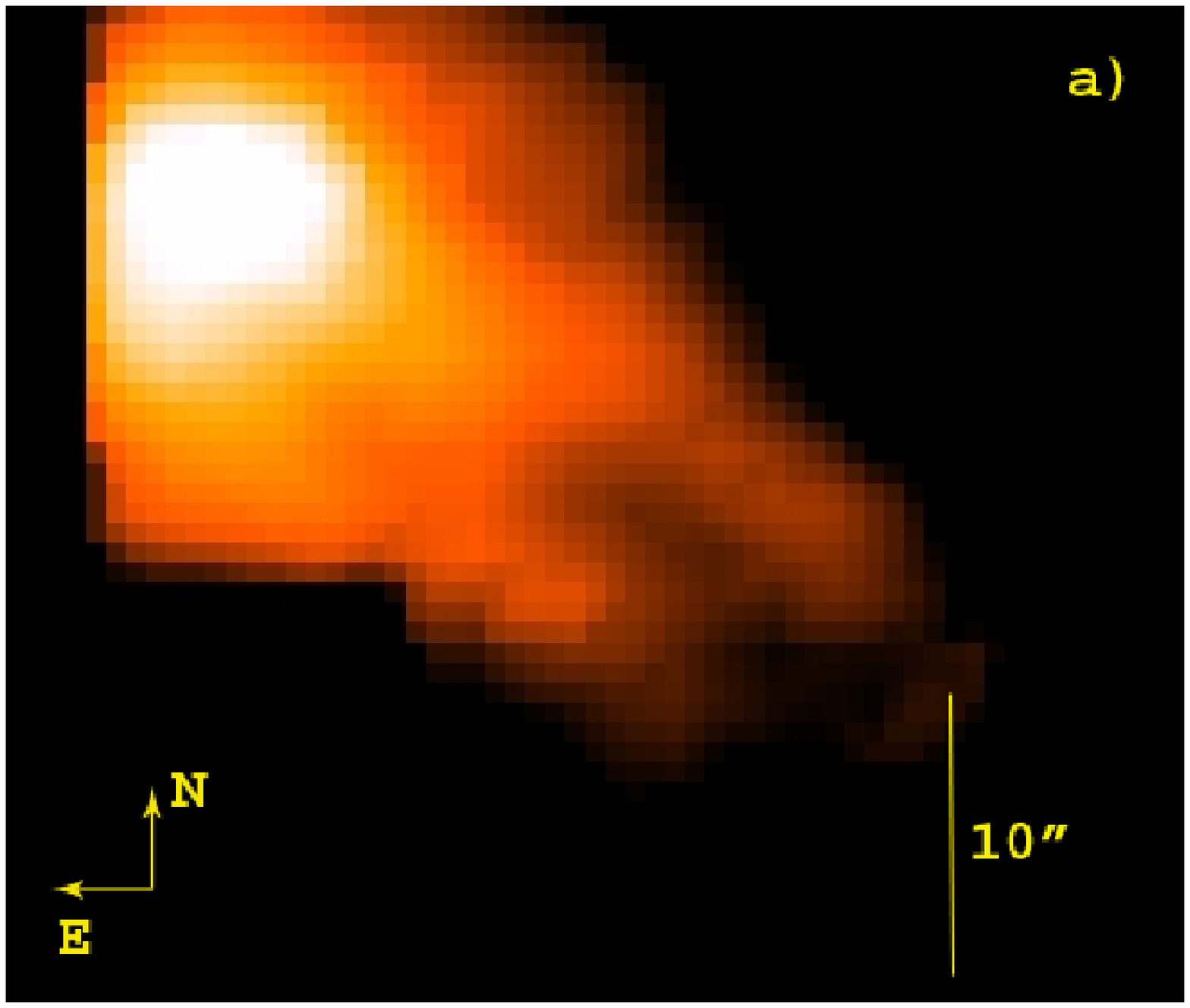}}
\hspace*{-0.04cm}\subfigure{\label{rotcen-a}\includegraphics[width=3.3cm]{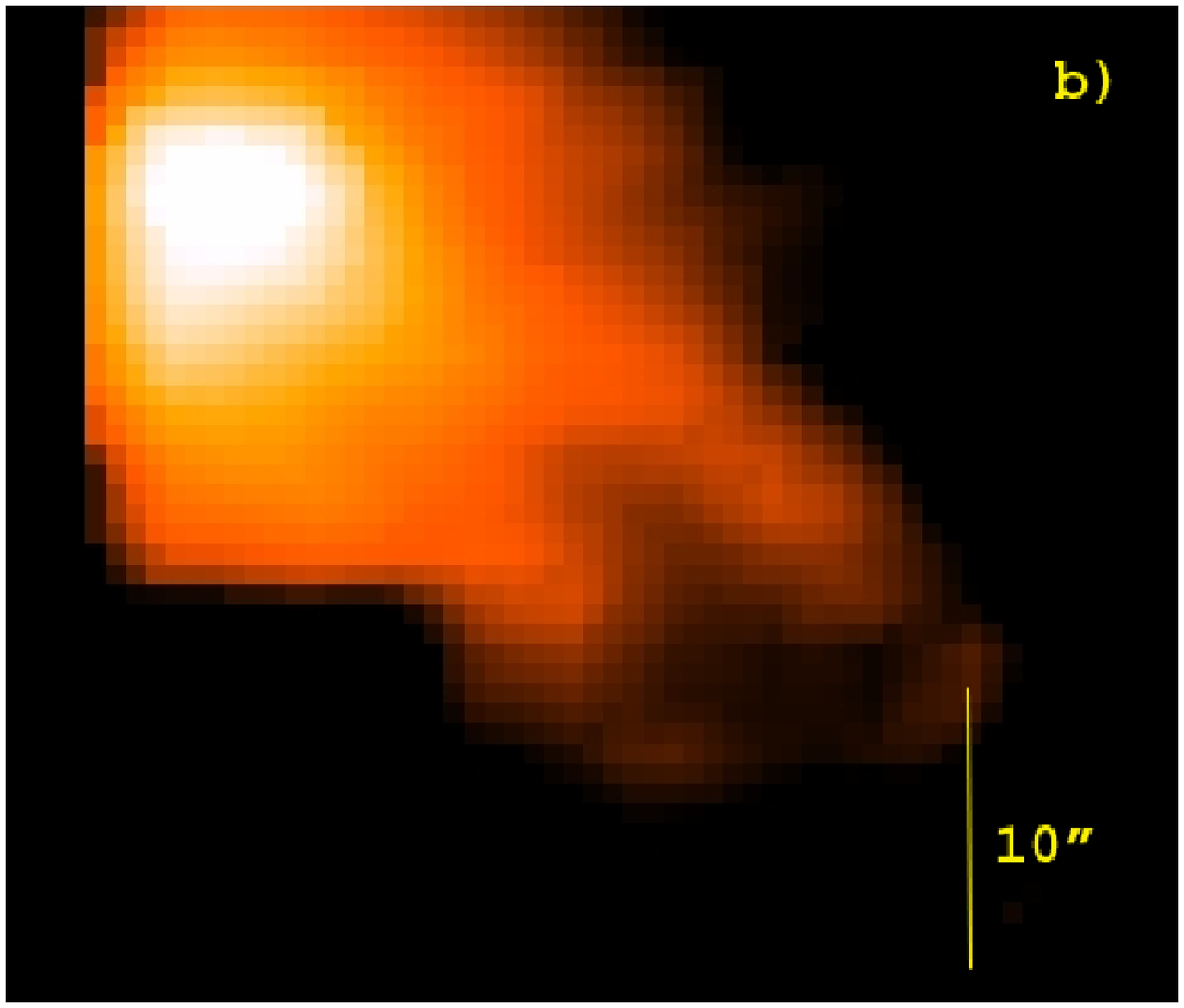}} 
   }}\\
\caption
{IIZw70~ maps in emission lines H$\alpha$$\lambda$6563\AA~(a) and [O {\sc iii}]$\lambda$5007\AA~(b). The
emission line images have been subtracted the continuum using images
using adjacent wavelengths obtained with PMAS at CAHA. North is up and east is left.}
\label{maps}
\end{figure}

In figure~\ref{maps} we show maps of IIZw70 in the flux for the lines
H$\alpha$$\lambda$6563 \AA~and [OIII]$\lambda$5007 \AA. The
original pixels have been resampled in order to smooth the emission
features. Emission line images are created by selecting appropriate
narrow band filters from the data cube. These narrow band images are
created in a wavelength range of $\pm$10 \AA~around the emission
wavelength and show emission line fluxes not corrected by
reddening. The continuum emission is subtracted using narrow-band
images in wavelengths adjacent to the emission lines.

\begin{figure}
\centerline{
   \includegraphics[bb=0 254 490 526,width=8cm,clip]{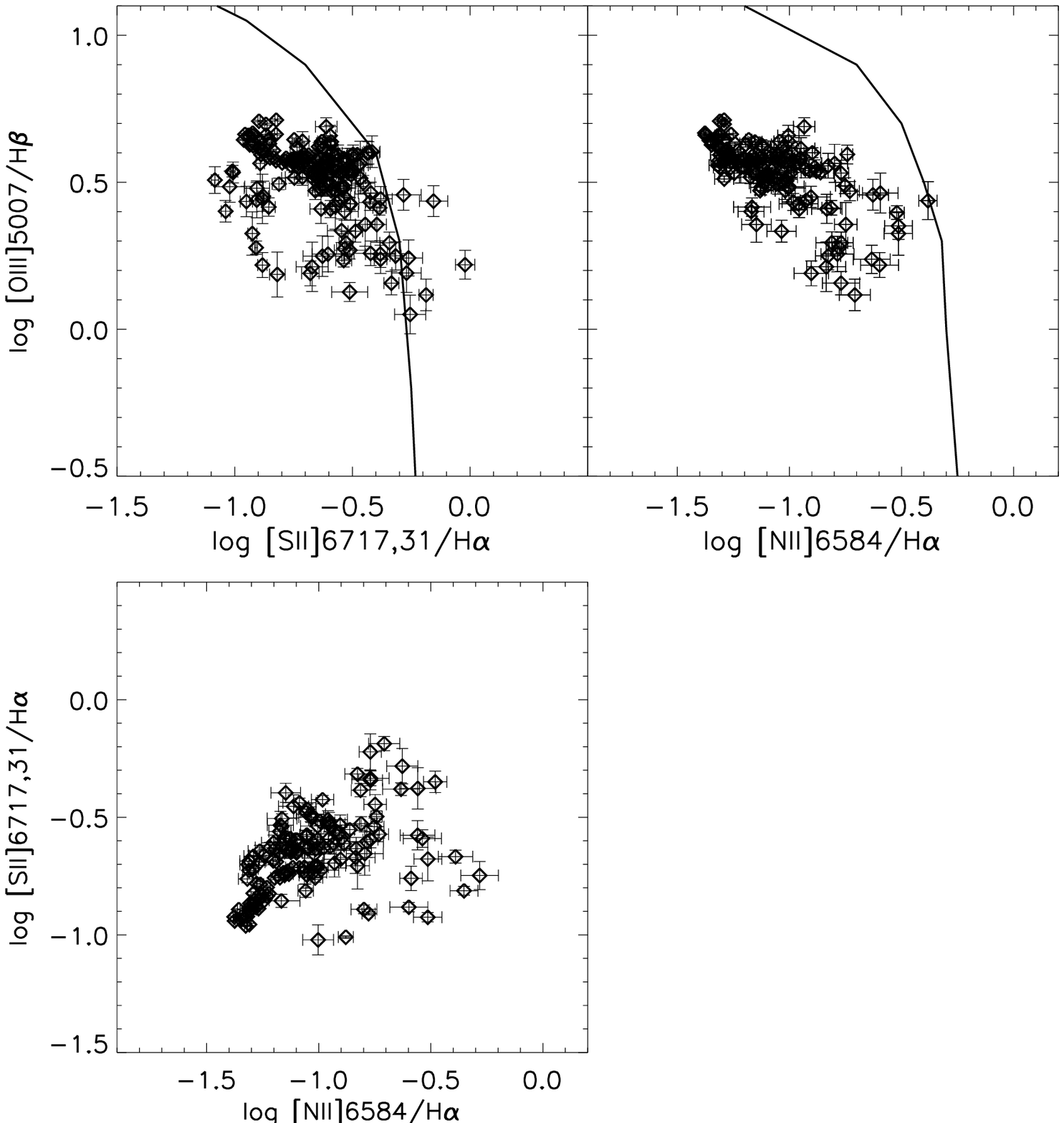}}
      \caption{Left panel shows the relation between [OIII]$\lambda$5007/H$\beta$ and [SII]$\lambda$$\lambda$6717,31/H$\alpha$ and right panel displays [OIII]$\lambda$5007/H$\beta$ line ratio as a function of [NII]$\lambda$6584/H$\alpha$; only line ratios with relative errors $\leq$ 30$\%$ are shown; the solid lines are the dividing lines between HII-regions galaxies (left) and AGNs (right) \cite{V87}.}
         \label{fig1}
   \end{figure}

Figure~\ref{fig1} shows the diagnostic diagrams,
[OIII]$\lambda$5007/H$\beta $ vs. [SII]$\lambda $6717,31/H$\alpha$
and [OIII]$\lambda $5007/H$\beta $ vs. [NII]$\lambda $6584/H$\alpha$
\cite{V87}, only for the fibers that present
diagnostic line-intensity ratios with relative errors $\leq$ 30$\%$.
Line fluxes were corrected for interstellar extinction using the value
of C(H$\beta$) for each fiber, applying the extinction law given by \cite{W58}. The solid lines show the classical separation between
ionization due to star formation (left) and active galactic nuclei
(AGN) (right). The boundary curve between AGNs and star forming regions
is taken from \cite{V87}.


The diagnostic diagrams give us information about the nature of the
ionizing source.  We can see that most of line-intensity ratios
measured for each fiber, at both diagrams, are located on the left
side of the solid curves. This suggests that for the galaxy IIZw70 the
ionising radiation is mainly related to the most massive youngest
stars.

We would like to stress that spectrophotometric studies of HII galaxies are
of paramount importance in order to derive their evolutionary
status. Besides, high-resolution spectroscopy of HII galaxies will help trace
the kinematics of the galaxy and provide valuable information on the
mechanisms triggering the star formation. A detailed analysis of these
data is going to be presented in a forthcoming paper.



\printindex
\end{document}